\documentclass[pdflatex,sn-mathphys-num,iicol]{sn-jnl}

\usepackage{graphicx}%
\usepackage{multirow}%
\usepackage{amsmath,amssymb,amsfonts}%
\usepackage{amsthm}%
\usepackage{mathrsfs}%
\usepackage[title]{appendix}%
\usepackage{xcolor}%
\usepackage{textcomp}%
\usepackage{manyfoot}%
\usepackage{booktabs}%
\usepackage{algorithm}%
\usepackage{algorithmicx}%
\usepackage{algpseudocode}%
\usepackage{listings}%
\usepackage{cleveref}%
\usepackage{physics}%

\DeclareMathOperator*\bigcircop{\bigcirc}

\theoremstyle{thmstyleone}%
\theoremstyle{thmstyletwo}%

\theoremstyle{thmstylethree}%

\begin{document}

\title[Quantum Deep Sets and Sequences]{Quantum Deep Sets and Sequences}


\author[1]{\fnm{Vladimir} \sur{Vargas-Calderón}}\email{vvargasc@dwavesys.com}
\affil[1]{\orgname{D-Wave Systems Inc.}, \orgaddress{\city{Burnaby}, \state{British Columbia}, \country{Canada}}}


\abstract{This paper introduces the quantum deep sets model, expanding the quantum machine learning tool-box by enabling the possibility of learning variadic functions using quantum systems.
  A couple of variants are presented for this model.
  The first one focuses on mapping sets to quantum systems through state vector averaging: each element of the set is mapped to a quantum state, and the quantum state of the set is the average of the corresponding quantum states of its elements.
  This approach allows the definition of a permutation-invariant variadic model.
  The second variant is useful for ordered sets, i.e., sequences, and relies on optimal coherification of tristochastic tensors that implement products of mixed states: each element of the set is mapped to a density matrix, and the quantum state of the set is the product of the corresponding density matrices of its elements.
  Such variant can be relevant in tasks such as natural language processing.
  The resulting quantum state in any of the variants is then processed to realise a function that solves a machine learning task such as classification, regression or density estimation.
  Through synthetic problem examples, the efficacy and versatility of quantum deep sets and sequences (QDSs) is demonstrated.}




\maketitle
\section{Introduction}\label{sec1}

Machine learning (ML) is usually concerned with modelling functions that look like $f:\mathbb{R}^d\to \mathcal{Y}$ in the case of supervised learning (such as classification and regression) and some unsupervised learning algorithms (such as clustering or generative modelling).
This paradigm restricts the kind of tasks that can be learnt using these ML algorithms.
Examples of important tasks not treatable with these ML models are point cloud classification and regression~\citep{pmlr-v139-goyal21a}, galaxy red-shift estimation~\citep{zaheer2017}, set anomaly detection~\citep{zaheer2017}, text concept set retrieval~\citep{alsaffar1999retrieval}, image tagging~\citep{Fu_Rui_2017}, among others.
The common factor between these tasks is that they can be posed as learning a function in a set space, instead of the customary vector space.
More specifically, these tasks require mapping a set of objects belonging to a space $\mathcal{S}$ to a target space $\mathcal{Y}$, formally defining the domain and codomain of the function to be modelled as $f:\mathcal{P}(\mathcal{S})\to\mathcal{Y}$, where $\mathcal{P}(\mathcal{S})$ is the power set of $\mathcal{S}$.

\citet{zaheer2017} introduced the Deep Sets model to learn functions that take the form $f(X) = h\left(\sum_{x\in X}g(x)\right)$, which are functions that map a set $X$ of objects $x$ to a given codomain $\mathcal{Y}$.
This map is done through the action of an embedding function $g$ and a projection function $h$.
The embedding function $g$ takes the objects $x$ into an embedding space; then, the sum over the embedded objects (which ensures permutation invariance) creates a new object in the embedding space containing information from all the objects in the original set $X$; finally, the projection function $h$ maps the resulting embedding of the set into the specified codomain $\mathcal{Y}$.

Deep Sets leverage neural networks to parameterise the functions $h$ and $g$, creating a powerful framework for processing sets.
The embedding function $g$ plays a crucial role by encoding complex data correlations into a representation vector space.
A representation of the set is obtained by adding the embedding vectors corresponding to the set's elements.

This work proposes using quantum physical systems to realise Deep Sets.
This is done either as a quantum-inspired method, where the embedding function $g$ maps data to the quantum state of some simulated quantum physical system and the projection function $h$ remains a neural network, or as a fully-quantum method, where $h$ is also realised as a physical operation on a quantum state representing a set of data points.
Let the reader be warned that the scaling of this proposal to system with more qubits is subject to the progress of data-encoding strategies~\citep{shin2023exponential,Wiersema2024herecomessun}, as usual strategies to encode data in the state of quantum circuits are known to limit the capabilities of variational quantum algorithms~\citep{li2022concentration, gil2024expressivity,thanasilp2024exponential,wang2025limitationsamplitudeencodingquantum}.

This paper introduces quantum deep sets in~\cref{sec:quantumdeepsets}, where the performance of this method in comparison to the classical deep sets algorithm is examined.
In~\cref{sec:qdseqs}, an extension to quantum deep sequences is discussed, addressing scenarios where the order of elements is crucial, and compare their performance against traditional long short-term memory (LSTM) networks in the context of a specific sequence classification problem.
Finally,~\cref{sec:conclusions} offers concluding remarks.


\section{Quantum Deep Sets}\label{sec:quantumdeepsets}

\citet{zaheer2017} introduced Deep Sets as a means to model functions of sets with domain $\mathcal{P}(\mathbb{R}^{d_\text{in}})$ and codomain $\mathbb{R}^{d_\text{out}}$.
They argued that these functions take the form
\begin{align}
    f(X) = h\left(\sum_{\vb*{x}\in X} g(\vb*{x})\right),\label{eq:originaldeepset}
\end{align}
where $X\in\mathcal{P}(\mathbb{R}^{d_\text{in}})$ is a set of vectors in $\mathbb{R}^{d_\text{in}}$.
The embedding and projection functions have signatures given by $g:\mathbb{R}^{d_\text{in}}\to\mathbb{R}^{d_\text{emb}}$ and $h:\mathbb{R}^{d_\text{emb}}\to\mathbb{R}^{d_\text{out}}$, respectively.

In fact, one can generalise the expression in~\cref{eq:originaldeepset} to
\begin{align}
    f(X) = h\left(\bigcircop_{\vb*{x}\in X}g(\vb*{x})\right),\label{eq:deepset_arbitrary_binary}
\end{align}
where $\circ$ stands for a binary operation that is both commutative and associative.
It is clear that the resulting function $f$ is still permutation invariant and variadic, which are the needed properties to define a function of sets.

For quantum deep sets, the embedding function will have a different (but equivalent) signature, i.e., $g:\mathbb{R}^{d_\text{in}}\to\mathcal{H}$, where $\mathcal{H}$ is the Hilbert space of some quantum physical system.
Here it becomes clear why using the original Deep Sets proposal of adding embedding vectors is not amenable to the quantum case, as the resulting embedding vector representing the set would not be a normalised quantum state.
For this reason, a different binary operation is needed.

A candidate binary operation is the convolution of normalised quantum states.
The result of quantum convolution between quantum states is a normalised quantum state, so the convolution operation is a valid commutative and associative binary operation that realises quantum deep sets.
However, convolution of quantum states is not physically realisable~\cite{lomont2003quantum}.
Approximate quantum convolution operations can be defined~\citep{Bistron2023,Korzekwa_2018,Aniello_2019,Aniello_2019_2} but they are, in general, not commutative nor associative.
Without these properties, the order in which the elements of the set are fed to the function $f$ does matter, and we obtain models for sequences, not for sets.
This will be exploited and explained in further detail in~\cref{sec:qdseqs}.
For now, we focus on physically realisable quantum deep set models, which can be realised by modelling $f$ as
\begin{align}
    f(X) = h(v(\{g(\vb*{x}):\vb*{x}\in X\})),
\end{align}
where $v$ is a variadic and permutation invariant function.
In this work, $v:\mathcal{P}(\mathcal{H})\to\mathcal{H}$ is defined as the $L_2$ average of quantum statevectors, i.e.,
\begin{align}
    \ket{v(X)} = \frac{\sum_{\vb*{x}\in X} \ket{g(\vb*{x})}}{\norm{\sum_{\vb*{x}\in X} \ket{g(\vb*{x})}}_2},\label{eq:average_of_quantum_states}
\end{align}
where it was made explicit that $v(X)$ and $g(\vb*{x})$ are quantum statevectors by using Dirac's notation.
Such a function $v$ can be realised in quantum circuits with algorithms such as linear combination of unitaries~\citep{childs2012hamiltonian,Chakraborty2024implementingany}.

The full quantum deep sets pipeline is thus specified by defining the embedding and projection functions $g$ and $h$.
\citet{zaheer2017} used three-layer fully connected neural networks as $h$ and $g$.
In the present work, variational parameters $\vb*{\theta}$ are introduced to define the embedding function as a parameterised quantum feature map~\citep{schuld2019qmlfeatures}
\begin{align}
\ket{g(\vb*{x},\vb*{\phi})} = U(\vb*{\theta}(\vb*{x},\vb*{\phi}))\ket{0}^{\otimes n},    \label{eq:emb_fun_hilbert}
\end{align}
where $U$ is parameterised as a general $SU(2^n)$ variational quantum gate acting on $n$-qubits~\citep{Wiersema2024herecomessun}.
{The parameters $\vb*{\theta}$ of the quantum feature map are obtained by using a neural network; more formally,} $\vb*{\phi}$ are parameters of a neural network that performs a map $\vb*{x}\mapsto \vb*{\theta}\in \mathbb{R}^{4^n - 1}$ that outputs $4^n - 1$ numbers that are identified as the decomposition angles of $U$ onto the generator basis $\{G_m\}_{m=1}^{4^n - 1}$ of the $\mathfrak{su}(2^n)$ Lie algebra, where this decomposition is specifically given by the Magnus expansion~\citep{magnus1954exponential,BLANES2009151}
\begin{align}
    U(\vb*{\theta}) = \exp\left(\sum_{m=1}^{4^n-1}\theta_m G_m\right).\label{eq:sun_expansion}
\end{align}
The decomposition of unitaries into the Pauli string generators of the $\mathfrak{su}(2^n)$ Lie algebra has shown better trainability properties than other common decompositions~\citep{Wiersema2024herecomessun}.
However, compilation of such unitaries remains a challenge for hardware implementations of the proposed method~\citep{ge2024quantumcircuitsynthesiscompilation}, especially as the number of qubits is increased.
Nonetheless, excellent results can be achieved with few qubits, as is shown at the end of this section.
{It is worth noting that one can encode the classical data directly as $\theta_m = \varphi x_i$, for some specified $m$ and $i$ chosen by the user, where $\varphi$ is a variational parameter analogous to those used in the variational circuit literature.
However, in this work, a neural network is employed to learn the mapping from classical data to the angles of an $SU(2^n)$ unitary, as it allows a much more flexible model.}

\begin{figure}
    \centering
    \includegraphics[scale = 1]{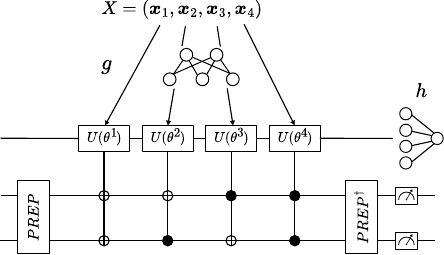}
    \caption{Depiction of a realisation of the quantum deep sets model. A neural network is used to transform each data point from the set $X$ into the angles of a $SU(2^n)$ gate, effectively performing the embedding function $g$. An LCU circuit is used {as an example} to prepare a superposition of the states prepared by each gate{, acting as the variadic and permutation invariant function $v$}.
    The resulting state is then fed into a projection neural network $h$ used to solve the ML task at hand.}
    \label{fig:LCU}
\end{figure}

First, the design of the projection function $h$ is clarified.
In a quantum circuit setup, $h$ can be realised in several ways, depending on the task for which one is trying to model a function of sets.
In general, $h$ involves post-processing measurement outputs of the form $\Tr[MV(\vb*{\varphi})\ketbra{v(X)}V^\dagger(\vb*{\varphi})]$, where $M$ is an element of some positive operator-valued measure, and $V(\vb*{\varphi})$ is a parameterised circuit~\citep{bermejo2022variationalquantumcontinuousoptimization}.
Even if it is possible to design $h$ as a quantum routine, there are many options to design post-processing routines.
Rather than exploring the various ways of implementing $h$ as a quantum routine, this work focuses on numerical experiments that treat quantum deep sets and sequences as a quantum-inspired algorithm, where it is sufficient to define $h$ as a full-forward neural network that ingests the quantum state described in~\cref{eq:average_of_quantum_states} and outputs a number for regression problems, or a class for classification problems.
A depiction of the quantum deep sets model is shown in~\cref{fig:LCU}.

In this context, where $g$ is realised by an $SU(2^n)$ variational quantum gate acting on $n$-qubits and $h$ is realised by a full-forward neural network, the quantum deep set model is benchmarked against the original Deep Sets model.
The benchmarking task proposed by~\citet{zaheer2017} consists of a regression task on a synthetic problem, where the entropy of the first dimension of multi-variate Gaussian distributions is to be estimated.
More specifically, consider a random $2\times 2$ covariance matrix $\Sigma$ that is used to draw a set $\mathcal{M}_i$ of $M_i$ two-dimensional points from the Gaussian distribution $\mathcal{N}(0, R(\alpha_i)\Sigma R^T(\alpha_i))$, where $\alpha_i$ is an angle uniformly drawn from $\mathcal{U}(0,\pi)$, and $R(\alpha)$ is the $2\times 2$ rotation matrix with angle $\alpha$.
The integer $M_i$ is uniformly drawn from $\mathcal{U}(300, 500)$.
$i$ is just an index to refer to a specific sample set.
Thus, the regression task specifically consists of, given the sample set $\mathcal{M}_i$, predict the entropy of the first dimension of $R(\alpha_i)\Sigma R^T(\alpha_i)$, i.e., $\frac{1}{2}(1+\log(2\pi\vb*{r}^T\Sigma^T\vb*{r}))$, where $\vb*{r} = (\cos\alpha_i, -\sin\alpha_i)^T$.

\begin{figure}[t]
    \centering
    \includegraphics[width=\columnwidth]{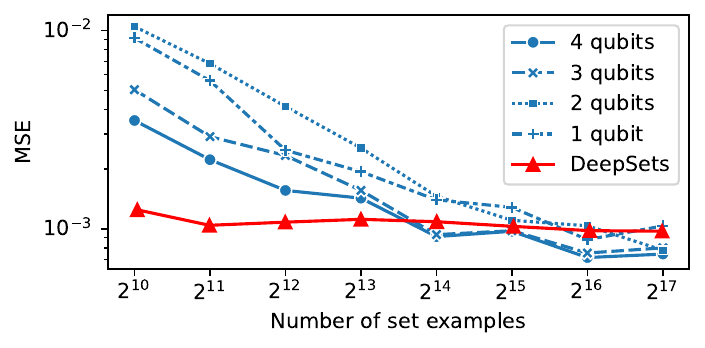}
    \caption{Mean squared error (MSE) as a function of the number of sample sets.
    The red line corresponds to a classical deep set model (data from Ref.~\citep{zaheer2017}).
    The blue curves correspond to quantum deep set models for different number of qubits $n$.
    In~\cref{eq:sun_expansion}, the neural network, which maps data points sampled from randomly rotated Gaussians (see main text) to the $SU(2^n)$ angles, has 8 hidden neurons.
    The three-layer neural network $h$ has 100 hidden neurons.
    MSE is calculated with respect to an independent test set consisting of sample sets obtained from $2^{10}$ angles $\alpha$.
    Lines are for guiding the eye.}
    \label{fig:deepsets}
\end{figure}

\citet{zaheer2017} used three-layer fully connected neural networks as $h$ and $g$.
Here, three-layer fully connected neural networks are used as well to realise $h$ and $g$, but the latter neural network is used to define the rotation angles of the $SU(2^n)$ gate defined in~\cref{eq:sun_expansion}.
\Cref{fig:deepsets} shows the performance of quantum deep sets in comparison to the classical deep sets model.
It can be seen that quantum deep sets are able to achieve a better performance than classical deep sets, though the number of samples to achieve such performance is larger.
Moreover, increasing the number of qubits for representing data improves the performance; however, this comes at an exponential runtime cost.


\section{Quantum Deep Sequences}\label{sec:qdseqs}
As previously mentioned, approximate quantum convolution operations could be used to define models given by~\cref{eq:deepset_arbitrary_binary}.
\citet{Bistron2023} showed that, through the process of coherification~\citep{Korzekwa_2018}, one can build binary operations on quantum states, namely, binary quantum channels that define products of quantum states, some of which can approximate quantum convolutions.
This requires yet another modification to the signature of the embedding function $g$.
Now, $g$ does not merely prepare pure states in a Hilbert space $\mathcal{H}$, but it prepares general quantum states in a density matrix space $\Omega_{2^n}$ of a system of $n$ qubits.
The prescription for building such a quantum channel for $n$ qubits requires a tristochastic tensor $T\in\mathbb{B}^{2^n\times 2^n\times 2^n}$.
The density matrix channel that we are interested in is defined as
\begin{align}
\begin{aligned}
    \circ_T:\Omega_{2^n}\times\Omega_{2^n} &\to\Omega_{2^n}\\
    (\rho, \sigma)&\mapsto \rho \circ_T \sigma = \Tr_{2}[V(\rho\otimes\sigma)V^\dagger],
\end{aligned}\label{eq:product_of_dm}
\end{align}
where $\rho$ and $\sigma$ density matrices, $\Tr_2$ is the partial trace over the second copy of $\Omega_{2^n}$, and $V$ is a unitary matrix whose components are given by $V_{kN+i,lN+j} = T_{klj}B_{il}^k$.
Here $\{B^k\}_{k+1}^N$ are, in general, $N=2^n$ $SU(N)$ matrices, the first of which can be fixed to be the identity.
Note that these matrices need not span $SU(N)$.
In fact, one can parameterise these matrices (as in~\cref{eq:sun_expansion}) and learn the parameters that make $\circ_T$ behave as close as possible to a commutative and associative binary operation.
The binary operation in~\cref{eq:product_of_dm} is a binary quantum channel that is trace preserving and completely positive~\citep{Bistron2023}.
{It is worth noting that~\cref{eq:product_of_dm} does not define the most general binary operation, as it restricts the dynamics of the two coupled subsystems with states $\rho$ and $\sigma$ to the unitary case.
A more general parameterisation of the Kraus representation of the binary quantum channel could be used in the future to expand the search space for such a binary operation.}

In general, these quantum channels are not commutative nor associative~\citep{Aniello_2019,Aniello_2019_2}.
A necessary but not sufficient condition for these properties is that the tristochastic permutation tensor $T$ induces these properties.
The interested reader is referred to the work by \citet{bistron2024continuous}, where details about these binary quantum channels are given.
For the purposes of this paper, it suffices to use these channels, defined in~\cref{eq:product_of_dm}, as a pooling operator in the embedding space $\Omega_{2^n}$.

The embedding function, similar to~\cref{eq:emb_fun_hilbert}, parameterises density matrices through~\citep{zfiyczkowski1998}
\begin{align}
\begin{aligned}
    g(\vb*{x},\vb*{\phi}_\theta, \vb*{\phi}_w) &= U(\vb*{\theta}(\vb*{x},\vb*{\phi}_\theta))\operatorname{diag}(\vb*{\lambda}(\vb*{w}(\vb*{x},\vb*{\phi}_w)))\\
    {}&\phantom{=.}{}U^\dagger(\vb*{\theta}(\vb*{x},\vb*{\phi}_\theta)),
\end{aligned}
    \label{eq:qfm_dm}
\end{align}
where $\vb*{\phi}_\theta$ are parameters of a neural network that performs the map $\vb*{x}\mapsto \vb*{\theta}\in\mathbb{R}^{4^n - 1}$, $U$ is given by~\cref{eq:sun_expansion}, and $\vb*{\phi}_w$ are parameters of another neural network that performs the map $\vb*{x}\mapsto\vb*{w}\in[0,1]^{2^n}$, whose output is used to construct the elements of the diagonal matrix $\operatorname{diag}(\vb*{\lambda})$ as follows

\begin{align}
    \begin{aligned}
        \lambda_1 &= 1 - w_1^{1/(N-1)}(\vb*{x},\vb*{\phi}_w),\\
        \lambda_k &= \left(1-w_k^{1/(N-k)}(\vb*{x},\vb*{\phi}_w)\right)\left(1-\sum_{i=1}^{k-1}\lambda_k\right).
    \end{aligned}\label{eq:lambdas_qfm}
\end{align}

\begin{figure}
    \centering
    \includegraphics[scale = 1]{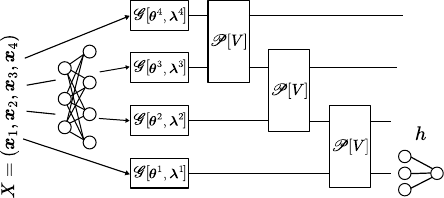}
    \caption{Depiction of the quantum deep sequences model. Each data point of the input set $X$ is passed through a neural network that predicts the parameters of a dynamical channel $\mathscr{G}$ that prepares a density matrix modelled under~\cref{eq:qfm_dm}.
    The binary quantum channel $\mathscr{P}$ that performs the product of two density matrices (cf.~\eqref{eq:product_of_dm}) is shown as a box that accepts two input density matrices and outputs on the bottom line the result of the product, after tracing out the degrees of freedom of the top line degrees of freedom. Finally, the resulting density matrix is passed to a full-forward neural network $h$ to solve the ML task at hand.}
    \label{fig:QDSeq}
\end{figure}

Diagramatically, the quantum deep sequences model is shown in~\cref{fig:QDSeq}.

\begin{figure}[t]
    \centering
    \includegraphics[width=\columnwidth]{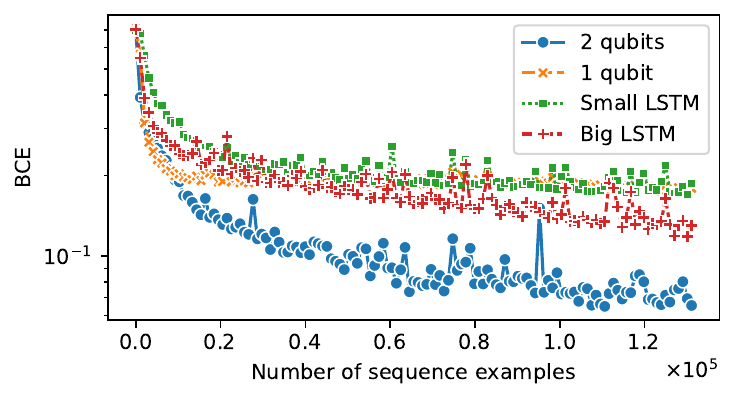}
    \caption{Binary cross-entropy (BCE) between the true class and the predicted class as a function of the number of training examples for sorted-in-increasing-order sequence classification.
    The number of trainable parameters for each model are 723, 653, 2053, and 1979 for the small LSTM, 1 qubit QDS, big LSTM, and 2 qubit QDS, respectively.
    BCE is calculated with respect to an independent test set of $2^{10}$ balanced sequences. Lines are for guiding the eye.}
    \label{fig:increasing_seq_classification}
\end{figure}

To showcase quantum deep sequence models, as a first learning task, consider the following binary classification problem.
A sequence of length $S_i$ of numbers is given; the quantum deep sequence model must classify it as either being sorted in increasing order or not.
$S_i$ is an integer sampled from $\mathcal{U}(10, 50)$. $i$ indexes the sequence.
The projection function $h$ has one output: $\hat{p}_i$ the probability that the sequence is classified as an increasing sequence.
\Cref{fig:increasing_seq_classification} shows the mean binary cross-entropy (BCE), defined as
\begin{align}
    \text{BCE}(\hat{p}, y) = -(y\log \hat{p} + (1-y)\log(1-\hat{p})),
\end{align}
as a function of the number of training sequences, which are balanced so that roughly 50\% of the samples are in increasing order.
$y$ is the true label of a particular sequence, which takes the value of 1 if it is ordered in increasing order, or 0 if it is not.

\Cref{fig:increasing_seq_classification} shows that for this simple sequence classification task, the quantum deep sequence model either matches or outperforms a similarly-sized long short-term memory (LSTM) model~\citep{hochreiter1997long}.

\section{Conclusion}\label{sec:conclusions}

This work introduced a quantum-inspired algorithm that implements models of variadic functions, i.e., functions of sets or sequences, through products of quantum states.
The algorithm consists of three steps.
The first one encodes data, corresponding to the elements of the set or sequence to be mapped, into the state of a quantum system.
The second one aggregates these quantum states into the quantum state of a single quantum system, which summarises the information of the set or sequence.
The final step maps the quantum state onto a scalar or vector that solves a regression or classification problem.
In this work, we explored a variation of this algorithm where the third step is done with a classical neural network.

The algorithm was tested on two problems: regression on sets and binary classification of sequences.
In both cases, the presented algorithm showed better performance than either the classical deep sets model or a classical sequence classification model (the LSTM).
The reason of such performance remains unexplored.
However, the models' performance improves with more representational power, i.e., with quantum system composed of more qubits, at the cost of exponential resources needed for training.
{Future work will need to focus on characterising the computational complexity and learning bounds of the presented algorithms.}
Moreover, it is expected from recent results on classical data encoding into quantum states, that performance starts to suffer degradation as more qubits are considered because of exponential concentration of the encoded quantum states{~\citep{thanasilp2024exponential}}.
Progress of data encoding will dictate the applicability of these algorithms to systems of high-dimensionality.

\backmatter

\bmhead{Acknowledgements}

I would like to thank Rafa{\l} Bistro{\'n}, Yudong Cao, Peter Johnson, Brian Dellabeta and Artem Strashko for comments on the version of this work that was presented at the Quantum Techniques in Machine Learning conference in 2024.
I would also like to thank Rafa{\l} Bistro{\'n}, Nicolas Parra-A. and Juan E. Ardila-García for enriching discussions and comments on the manuscript.
Particularly, Juan E. Ardila-García also helped me with designing and drawing~\cref{fig:LCU,fig:QDSeq}.

\bmhead{Competing Interests}
The authors declare the following competing interests: V.V.-C. was employed by Zapata Computing Inc. during the majority of the development of this work.









\bibliography{sn-bibliography}

\end{document}